\documentclass[aps,prl,showpacs,nofootinbib,floats,floatfix,preprintnumbers
, groupedaddress, twocolumn]{revtex4}
\usepackage{comment}
\usepackage{graphicx}
\usepackage{amsmath}
\usepackage{float}
\usepackage{subfigure}
\usepackage[usenames]{color}
\usepackage{amssymb}
\usepackage{epstopdf}
\newcommand{\bea}{\begin{aligned}}
\newcommand{\eea}{\end{aligned}}
\newcommand{\be}{\begin{equation}}
\newcommand{\ee}{\end{equation}}

\newcommand{\nno}{\nonumber}
\newcommand{\bse}{\begin{subequations}}
\newcommand{\ese}{\end{subequations}}

\usepackage{hyperref}
\hypersetup{
     colorlinks   = true,
     citecolor    = red,
     linkcolor    = blue,
     urlcolor     = blue,
}

\newcommand{\bmm}{\begin{multline}}
\newcommand{\emm}{\end{multline}}

\numberwithin{equation}{section}
\begin{document}
\title{Gravitational reheating	}
\author{Md Riajul Haque\footnote{riaju176121018@iitg.ac.in, riajul.haque@physics.iitm.ac.in}}
\affiliation{Department of Physics, 
Indian Institute of Technology, Guwahati, India}
\affiliation{Centre for Strings, Gravitation and Cosmology, Department of Physics, Indian Institute of Technology
Madras, Chennai 600036, India}
\author{Debaprasad Maity, \footnote{debu@iitg.ac.in}}
\affiliation{Department of Physics, 
Indian Institute of Technology, Guwahati, India}
\pagenumbering{arabic}
\renewcommand{\thesection}{\arabic{section}}

\begin{abstract}
Our present understanding of the reheating phase is incomplete due to a lack of observations. Apart from its cosmological
implications, the reheating should play a vital role in particle physics and inflation model building. Conventionally reheating dynamics are modeled by invoking arbitrary coupling among the inflaton and daughter fields. Such an approach lacks robust cosmological predictions due to its arbitrary couplings and is difficult to verify through observation. In this paper, we propose a minimal reheating scenario where the inflaton is coupled with all the daughter fields only gravitationally. Besides being successful in reheating the Universe, the scenario offers a strong cosmological prediction of the primordial gravitational wave spectrum and discards a large number of possible models of dark matter and inflation that are otherwise consistent with Planck.

%

\end{abstract}

\maketitle
\newpage

\begin{figure}[t]
	{\includegraphics[width=8.2cm,height=5.3cm]{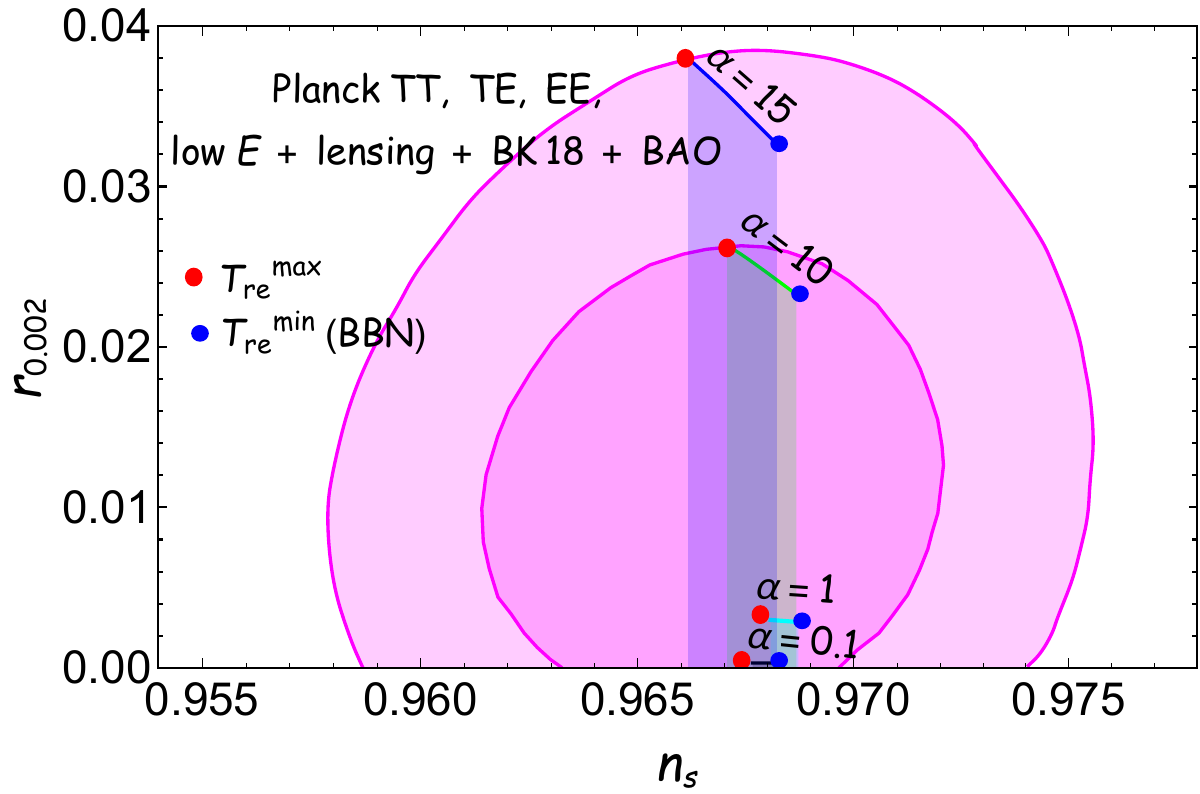}}
	\caption{Compare our result for $\alpha$-attractor model with different values of $\alpha$ in the $(n_s, r)$ plane based on the observational $68\%$ and $95\%$ CL constrain from the combined data of recent BICEP/Keck and Planck.}
	\label{Bicep/keckconstrain}
\end{figure}

{\bf Introduction:}
Reheating is a natural physical phenomenon after inflation when dark matter (DM) and all standard model (SM) particles may be produced. In the simplest scenario, when a single scalar field drives inflation, shift symmetry is expected to play an important role in the nature of coupling among inflaton and any of the other fields, and this symmetry must naturally suppress it. However, all fields are naturally coupled to gravity through s-channel graviton $(h_{\mu\nu})$ exchange interaction, $\mathcal (1/M_P^2)h_{\mu\nu}T^{\mu\nu}$, and when the energy scale of any physical processes such as reheating is as large as $\sim 10^{15}$ GeV, gravity mediated decay process may be strong and sufficient to reheat the Universe. Where $T^{\mu\nu}$ corresponds to the energy-momentum tensor of all the fundamental fields, this is the possibility we will explore in this paper. We will name it gravitational reheating (GRE).
In this phase, DM mass is the only free parameter except, of course, the inflationary parameters. We will see how such less freedom naturally makes GRE a model-independent mechanism as compared to reheating scenarios discussed so far in the literature \cite{Giudice:2000ex,Haque:2020zco,Dai:2014jja,Drewes:2017fmn}. All the massless decay products from inflaton will be collectively called radiation, and massive ones are DM. Given the present state of the Universe, GRE turned out to be consistent with a very limited class of inflation models and a narrow range of DM masses. GRE is insensitive to any new physics in the radiation and DM sector. However, if DM couples with the radiation bath, gravitational production sets the maximum limit on the DM mass \cite{Haque:2021mab}.
It is the s-channel graviton exchange process through which inflaton converts its energy to radiation and DM during reheating. Gravitaton exchange processes between radiation bath and DM will be ignored due to its sub-dominant contribution (see detailed study in \cite{Garny:2015sjg,Haque:2021mab,Clery:2021bwz}). The dynamical equations for GRE are \cite{Giudice:2000ex,Haque:2020zco}
\begin{eqnarray}
\label{Boltzmann}
&&\dot{\rho}_\phi +3H(1+\omega_\phi)\rho_\phi+\Gamma^T_{\phi}\rho_\phi(1+\omega_\phi)=0\,,\nonumber\\
&&\dot{\rho}_R+4H\rho_R-\Gamma_{\phi\phi\to RR}^{Rad}\,\rho_\phi(1+\omega_\phi)=0\,,\\
&&\dot{n}_Y+3Hn_Y-\frac{\Gamma_{\phi\phi\to YY}^{DM}}{m_\phi}\,\rho_\phi(1+\omega_\phi)=0\,,\nonumber
\end{eqnarray}
where, $(\rho_\phi,\rho_R,n_Y)$ are inflaton energy density, radiation energy density and dark matter number density respectively. The total inflation decay width is $\Gamma^T_{\phi}=\Gamma_{\phi\phi\to RR}^{Rad}+\Gamma_{\phi\phi\to YY}^{DM}$.
The gravitational decay widths of inflaton to fundamental fields are \cite{Donoghue:1994dn,Choi:1994ax,Holstein:2006bh,Mambrini:2021zpp,Barman:2021ugy},
\begin{eqnarray}\label{decay}
&&\Gamma_{\phi\phi\to SS}=\frac{\rho_\phi\,m_\phi}{1024\,\pi\,M_p^4}\,\left(\,1+\frac{m_S^2}{2\,m_\phi^2}\,\right)\sqrt{1-\frac{m_S^2}{m_\phi^2}}\,,\nonumber\\
&&\Gamma_{\phi\phi\to ff}=\frac{\rho_\phi\,m_f^2}{4096\pi\,M_p^4m_\phi}\left(1-\frac{m_f^2}{m_\phi^2}\right)^{\frac{3}{2}},\\
&&\Gamma_{\phi\phi\to XX}=\frac{\rho_\phi\,m_\phi}{32768\,\pi\,M_p^4}\,\left(4+4\frac{m_X^2}{m_\phi^2}+19\frac{m_X^4}{\,m_\phi^4}\right)\sqrt{1-\frac{m_X^2}{m_\phi^2}}.\nonumber
\end{eqnarray}
The symbols $(R,Y)$ represent scalar ($S$), fermion ($f$), and vector particles ($X$). Pauli spin blocking renders inflaton to fermion decay width proportional to the fermion mass $m_f$. This immediately indicates $\Gamma^{Rad}_{\phi\phi\to ff}\ll \Gamma^{Rad}_{\phi\phi\to SS},\Gamma^{Rad}_{\phi\phi\to XX}$, as the mass of the radiation constituents is very small compare to the inflaton mass (here we have taken the radiation particles as massless). Hence, we ignore the fermionic contribution in radiation baths throughout.  Consequently, $\Gamma^{Rad}_{\phi\phi\to RR} = \Gamma^{Rad}_{\phi\phi\to SS}+\Gamma^{Rad}_{\phi\phi\to XX} = (1+\gamma)\Gamma^{Rad}_{\phi\phi\to SS} $, with $\gamma =1/8$. For DM, we analyze individual species, and the mass of the DM can not exceed the inflaton mass due to kinematical reasons. Massless graviton can also be part of the radiation bath through s-channel production, whose decay width will be suppressed due to its tensorial structure like the electromagnetic field. We ignore it in our analysis throughout. 
\\
{\bf Model of inflation:}
To better understand the mechanism, along with the model-independent consideration, we also consider $\alpha$-attractor model with inflation potential \cite{Kallosh:2013hoa,Kallosh:2013yoa},
\be \label{alpha-attractor}
V(\phi)=\Lambda^4\,\left[1-e^{-\sqrt{\frac{2}{3\,\alpha}}\phi/M_p}\right]^{2n}
\ee
Where, $(\alpha, n, \Lambda)$ are free parameters. 
In this model, the inflationary observables assume remarkably simple form, $1-n_s\simeq {2}/{N_k},\, r\simeq {12\,\alpha}/{N_k^2}
$ \cite{Ellis:2013nxa}.
%
%
Inflationary e-folding, $N_k$ is defined for a CMB scale of interest $k$ which crossed the Hubble radius near the beginning of inflation. After inflation ends, inflaton undergoes damped oscillation due to decay around the minimum where potential assumes power-law form, $
V(\phi)=\lambda\,\phi^{2n} $, with  $\lambda=\Lambda^4 \left({2}/{(3\alpha M_p^2)}\right)^n$. 
In order to describe reheating dynamics, we assume the equation of state (EoS) of the inflaton averaging over oscillation to be $\omega_\phi\simeq (n-1)/(n+1)
$,
and effective mass of inflaton $m_\phi$ \cite{Drewes:2017fmn} in terms of energy density $\rho_{\phi}$ as
\be \label{effective mass}
m_\phi=
\sqrt{\frac{2(1+\omega_\phi)\,(1+3\omega_\phi)}{(1-\omega_\phi)^2}}\,\lambda^{\frac{1-\omega_\phi}{2\,(1+\omega_\phi)}}
\rho_\phi^{\frac{\omega_\phi}{1+\omega_\phi}}
\ee\\
{\bf Computing reheating parameters:}
In order to calculate quantities during reheating namely, reheating e-folding number $(N_{re})$, reheating temperature $(T_{re})$, and maximum radiation temperature $(T_{max})$, we evaluate Eq.\ref{Boltzmann} for radiation,  
\be
d\,(\rho_R\,A^4)=\Gamma_{\phi\phi\to RR}^{Rad}\,\rho_{\phi}\,(1+\omega_\phi)\,\frac{A^3\,dA}{H}
\ee
where, $A=a/a_{end}$ is normalized scale factor. Suffix $"end"$ corresponds to the end of inflation. The production of radiation will depend on the inflaton energy density only, and hence, maximum production occurs at the beginning of reheating. During this early stage inflaton is naturally the dominating component. Neglecting decay term, therefore, $\rho_{\phi}$ approximately evolves as $
\rho_\phi=\rho_\phi^{end}\,A^{-3\,(\,1+\omega_\phi\,)}\,,
$
where, $\rho_\phi^{end}=3\,M_p^2\,H_{end}^2$ denotes the inflaton energy density at the end of inflation. Consequently the Hubble parameter becomes, 
\be \label{Hubble}
H= \frac {\Lambda^2}{\sqrt{2} M_p}\left(\frac {2n}{2n + \sqrt{3\alpha} }\right)^n A^{-\frac{3}{2}(1+\omega_\phi)} =H_{end} A^{-\frac{3}{2}(1+\omega_\phi)}. \nno
\ee
 Using these the dynamical equation for the comoving radiation energy density transforms into,
\be \label{comovingrad}
d\,(\rho_R\,A^4)=3M_p^2 H_{end}\Gamma_{\phi\phi \to RR}^{Rad}(1+\omega_\phi) A^{\frac{3}{2}(1-\omega_\phi)}dA . \nno
\ee
With this we now calculate $(N_{re}, T_{re}, T_{max})$. 
Considering the massless limit of the radiation constituents 
%
and using Eq.\ref{effective mass} into, Eq.\ref{comovingrad}, we find
\be \label{radiationdensity}
\rho_R=\frac{9\,(1+\gamma)\,H_{end}^3\,m_\phi^{end}\,(\,1+\omega_\phi\,)}{512\,\pi\,(1+15\omega_\phi)\,A^4}\left(1-A^{-\frac{1+15\,\omega_\phi}{2}}\right)
\ee
The above equation suggests that radiation production quickly happens at the beginning of reheating for large inflaton energy density and then freezes out.
Competition between production and background expansion leads to a peak $T_{max}$ in the radiation temperature, which is expressed as, 
\be \label{tmax}
\left(T_{max}\right)^4=\frac{9(1+\gamma)H_{end}^3 m_\phi^{end}(1+\omega_\phi)}{512\beta \pi(1+15\omega_\phi)A_{max}^4}\left(1-A_{max}^{-\frac{1+15\omega_\phi}{2}}\right)
\ee
Here, $\beta={\pi^2 g_*^{re}}/{30}$ and $g_*^{re}$ denotes the effective number of degrees of freedom associated with the radiation bath at the point of reheating. Where, $A_{max}=((9+15\omega_\phi)/{8})^{\frac{2}{1+15\,\omega_\phi}}$.
The end of reheating is defined at the point where $\rho_\phi=\rho_R$ as long as it satisfies BBN temperature bound. It turns out that when $\omega_{\phi} <1/3$, the above condition is equivalent to $H \simeq \Gamma^{Rad}_{{\phi\phi \to RR}}$, which may not necessarily be true for $\omega_{\phi} > 1/3$. This is because the inflaton dilutes itself much faster than the radiation due to expansion. Thus even the condition $H \simeq \Gamma^{Rad}_{{\phi\phi \to RR}}$ is not satisfied reheating condition $\rho_\phi=\rho_R$ is achievable. Accordingly, Eq.\ref{radiationdensity} with the condition of reheating end $\rho_\phi=\rho_R$, one can obtain the reheating e-folding number $N_{re}$ as,
\be \label{Nre}
N_{re}=\frac{1}{3\omega_\phi-1} \ln \left(\frac{512\,\pi\,\,M_p^2\,(1+15\,\omega_\phi)}{3\,(1+\gamma)H_{end}\,m_\phi^{end}\,(1+\omega_\phi)}\right),
\ee
By using the above equation (Eq.\ref{Nre}) one immediately computes the reheating temperature as,
\be\label{Tre}
T_{re}=\left(\frac{9\,(1+\gamma)\,H_{end}^3\,m_\phi^{end}\,(\,1+\omega_\phi\,)}{512\,\beta\,\pi\,(1+15\omega_\phi)}\,e^{-4\,N_{re}}\right)^{1/4}
\ee
Furthermore,
entropy conservation from the reheating end to present time gives an additional important relation between $(T_{re}, N_k)$ as \cite{Dai:2014jja}
\be \label{entropy-conservation}
T_{re}=\left(\frac{43}{11\,g_*^{re}}\right)^{1/3}\,\left(\frac{a_0\,H_{end}}{k}\right)\,e^{-(N_k+N_{re})}\,T_0\,,
\ee
Where, the use has been made of the relation $a_k H_k = a_0 H_0$ for $k$ being CMB pivot scale, $k/a_0=0.05~\mbox{Mpc}^{-1}$. $T_0= 2.725$ K is the present CMB temperature.\\
\underline{\it Model independent constraints:} Combining both Eqns. \ref{Tre} and \ref{entropy-conservation}, the reheating temperature $T_{re}$ turns out to be the only function of inflationary parameters, $(\omega_{\phi},\, H_{end},\, m^{end}_{\phi})$. Here, we first discuss the generic bounds on de Sitter type inflation without specifying any particular model. Using the following approximate relation $m_{\phi}^{end} \simeq \sqrt{{(1+\omega_\phi)(4+12\omega_\phi)}/{(1-\omega_\phi)^2}}\,H_{end}$ (under the assumption $\phi_{end}\sim M_p$), one immediately gets $\omega_{\phi}$ within $(0.60,\,0.99)$ and $H_{end}$ within $( 1\times10^{9},5\times 10^{13})$ GeV. This narrow and closed bound are derived form the minimum reheating temperature set by BBN as $T_{re}^{min} = T_{BBN}\sim 10^{-2}$ GeV \cite{Kawasaki:2000en,Hannestad:2004px} and maximum possible value of the de Sitter Hubble scale at the end of inflation, $H^{max}_{end} \simeq \pi M_p \sqrt{r A_s/2}$ calculated at upper limit on $r=0.036$ \cite{BICEP:2021xfz} (see Fig.\ref{Bicep/keckconstrain}). Using these bounds GRE predicts reheating temperature to be $T_{re} \lesssim 10^{8}$ GeV. 
Furthermore, using Eq.\ref{entropy-conservation} we found that inflationary e-folding number $N_k$ has to be within a very narrow range $(62, 63)$. {\it Therefore, to have successful GRE, viable de Sitter inflation models will be those, which give $N_k \sim (62,63)$, stiff reheating EoS $\omega_\phi>1/3$ and predict reasonably low values of $T_{re}$}. This is indeed the case as we will discuss for $\alpha$-attractor.\\ In addition, we can also consider quantum vacuum production of all
the fields, including radiation (massless particles), due to the sudden transition from inflationary to reheating phase where the adiabaticity may be violated \cite{Ford:1986sy,Artymowski:2017pua}. The radiation energy density due to this quantum vacuum production is estimated as $\rho_R^{v}\simeq \frac{9(1+\omega_\phi)^2H_{end}^4}{128\pi^2\,A^{-4}}$. To derive this equation, the transition time scale between the de-sitter and the deaccelerating Universe is estimated as $x_0<<1$ and $log(1/x_0)\sim1$ \cite{Ford:1986sy,Artymowski:2017pua}. Including this quantum vacuum production with our s-channel scattering does not affect our prediction much; we ignore this quantum vacuum production throughout our analysis. However, including this effect the lower limit on the $\omega_\phi$ sifted $0.60\to0.59$.\\ 
\underline{\it Model dependent constraints:} Recent BICEP/Keck \cite{BICEP:2021xfz} data, in combination with Planck\cite{Planck:2018jri}, impose a new constraints on the scalar spectral index as well as on tensor-to-scalar ratio $r$. In recent studies Ref.\cite{Ellis:2021kad}, they have shown if one took $\alpha-$ attractor model (E-model) with matter like reheating by considering a non-gravitational coupling between inflaton and radiation field, $\alpha<26$ to satisfy the 2 $\sigma$ bound on $n_s-r$ from the combined data of BICEP/Keck and Planck. However, in our case, the situation is completely different. For GRE taking $\omega_\phi=0$ (matter like reheating) radiation-dominated era is not achievable. Moreover, for GRE if one assumes the $\alpha-$ attractor model (E-model) in order to reheat the Universe successfully, $\omega_\phi$ must be lies above $(0.65,\,0.65,\,0.67,\,0.68)$ for $\alpha=(0.1,\,1,\,10,\,15)$ respectively. Since there is no non-gravitational coupling in our analysis, there is a one-to-one correspondence between the parameters $(n_s,\,r,\,\omega_\phi,\, T_{re})$ once we fixed $\alpha$. This implies once $\alpha$ is fixed, for a particular value of $\omega_\phi$, we only have a specified value of $T_{re},\,n_s$ and $r$. In Fig.\ref{Bicep/keckconstrain}, we have shown where GRE lies in the $n_s-r$ plane with the latest available combined data from BICEP/Keck and Planck. An interesting finding of this analysis is that only $\alpha\leq15$ the results lie within the $95\%$ C.L and $\alpha\leq10$ at the $68\%$ C.L. We can say that this GRE scenario put tighter constraints on $\alpha$ in comparison with the case described in Ref.\cite{Ellis:2021kad} to be consistent with the recent BICEP/Keck and Planck data. Using the Big Bang Nucleosynthesis (BBN) constraints on the primordial gravitational wave and assuming GRE as a reheating background, the restriction on the upper limit of $\alpha$ is even more, tighter, which we have discussed later. One important point we want to point out is that we ignore further lower values of $\alpha$, $\alpha<0.1$ and the main reason behind this consideration is that for lower values of $\alpha$ ($\alpha<<1$), the self-resonance of the inflaton field is effective; we can not ignore it. However, for $\alpha\geq 1/6$, self-resonance is not important, and inflaton condensate remains intact \cite{Lozanov:2017hjm}. \\
\begin{figure}[t]
	{\includegraphics[width=8.7cm,height=3.3cm]{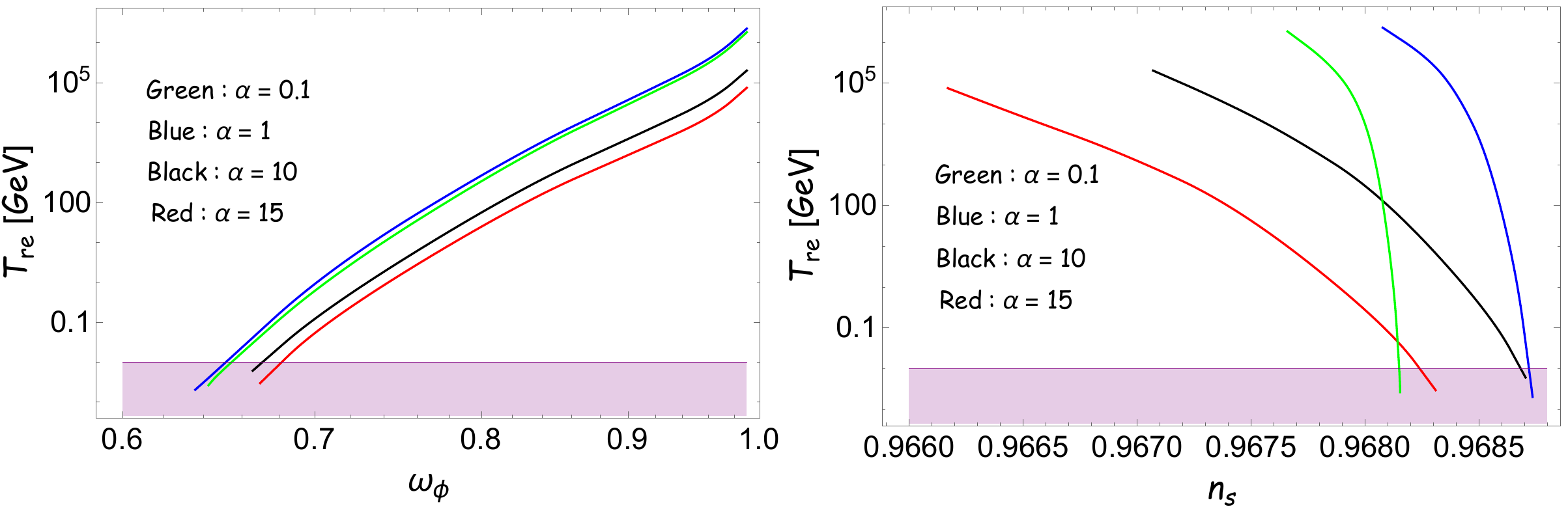}}
	\caption{{\bf Left panel:} Variation of $T_{re}$ as a function of $\omega_\phi$ with $\alpha-$ attractor model for four different sample values of $\alpha=(0.1,\,1,\,10,\,15)$. The purple region below $10^{-2}$ GeV is forbidden from BBN bound. 
{\bf Right panel:} Variation of $T_{re}$ as a function of $n_s$.} 
	\label{Reheatingtemp}
\end{figure} 
Taking four sample values of $\alpha=(0.1,\,1,\,10,\,15)$, the bound on the model parameters are found to be $\{(200.0 \geq n \geq 4.8),\,(200.0 \geq n \geq 4.8),\,(200.0 \geq n \geq 5.2),\,(200.0 \geq n \geq 5.3)\}$ and $\{(0.9677 \leq n_s \leq 0.9682),\,(0.9681 \leq n_s \leq 0.9687),(0.9671 \leq n_s \leq 0.9687),\,(0.9662 \leq n_s \leq 0.9682)\}$ respectively . The bounds are well within the $2\,\sigma$ range of $n_s$ (95 $\%$ CL, Planck TT,TE,EE+lowE+lensing) from Planck \cite{Planck:2018jri}. Moreover, those bounds in terms of reheating temperature turn out to be $\{ (10^{-2}\, \mbox{GeV} \leq T_{re} \leq 2\times10^6\,\mbox{GeV}),\,(10^{-2}\, \mbox{GeV} \leq T_{re} \leq 2\times10^6\, \mbox{GeV}),\,(10^{-2}\, \mbox{GeV} \leq T_{re} \leq 2\times10^5 \mbox{GeV}),\,(10^{-2} \mbox{GeV} \leq T_{re} \leq 7\times10^4 \,\mbox{GeV})$ for $\alpha=(0.1,\,1,\,10,\,15)$ accordingly and can be decoded from Fig.\ref{Reheatingtemp}. In all cases, the BBN energy scale sets the lower bound, and the upper bound is set by the kination equation of state $\omega_\phi \sim 1$.
Finally, model together with GRE as a background predicts $N_k$ within $\{(61.6,62.5),\,(62.5,63.3),\,(64.6,64.9)\,(62.5,63.3)\}$ for $\alpha=(0.1,\,1,\,10)$ and around $65.3$ for $\alpha=15$. Since the results for $\alpha=0.1$ and 1 are almost identical, in
our following discussions will consider only two sample values of $\alpha=(1,\,10)$ with all the above bounds to find the DM mass.
\\

{\bf Non-gravitational couplings: where do GRE lies?}
\begin{figure}
 	\begin{center}
\includegraphics[width=8.8cm,height=3cm]{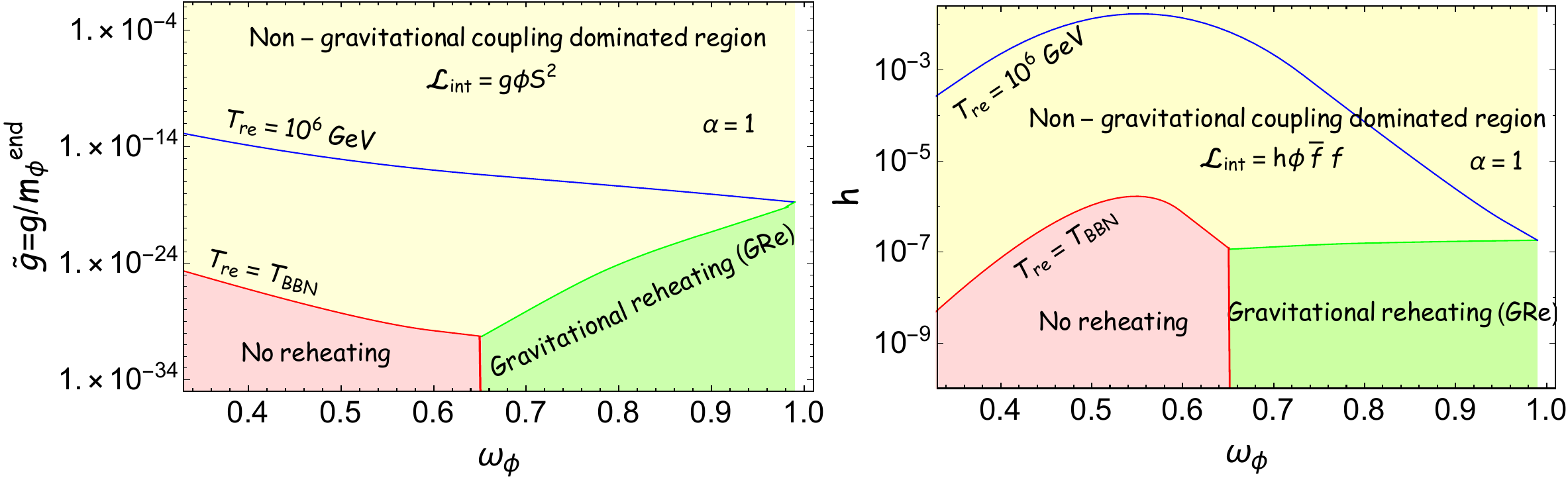}
\caption{\scriptsize 
Variation of $\tilde{g}$ and $h$ as a function of $\omega_\phi$. The yellow and green shaded regions show the explicit coupling dominated and purely GRE processes. In addition, the light-red shaded region corresponds to the no reheating regime for not satisfying the BBN energy scale. } 
 		\label{bound}
 	\end{center}
 \end{figure}
Besides gravitational interaction, we can always consider a non-gravitational interaction between inflaton and SM. To explain in which coupling parameter space our GRE scenario is sufficient to explain the present Universe, we assume two standard non-gravitational couplings between the inflaton SM sector: 1) The inflaton field coupled with the SM scalar via interaction $ \mathcal{L}_{int} = g\phi S^2$.  2) Assuming an interaction between the inflaton and SM fermions of the form $\mathcal{L}_{int} = h\phi\bar{f}f$. For the details of the calculation of how we can get results presented in Fig. \ref{bound}, follow the reference \cite{Haque:2023yra}. 
 From Fig.\ref{bound}, one important conclusion we arrive at is that GRE scenario plays the leading role in an extensive range of coupling parameters. 
As an example, for $\omega_\phi\sim1$, the coupling parameters in the limit of $h<10^{-7}$ and $\tilde{g}=g/m_\phi^{end}<10^{-19}$, GRE scenario works fine. \\
{\bf DM phenomenology:}
In particle physics, DM is still an ill-understood subject. Experimental direct detection proves to be challenging due to its unknown but tiny interaction with the nucleons. However, if the interaction is only gravitational, which is explicitly known, we may need to go beyond the conventional methods of detecting it. Planckian interacting dark matter has recently gained interest in the literature \cite{Mambrini:2021zpp,Barman:2021ugy}. 
In our GRE scenario, similar to radiation, DM is also coupled with inflaton suppressed by Planck mass. Therefore, DM mass $m_Y$ is the only free parameter. Interestingly such scenarios naturally fix the DM mass through its abundance and inflaton model under consideration. Dynamics of DM is governed by [see, for instance, the last expression of Eq.(\ref{Boltzmann})]
 \be \label{DMcomoving}
 d(n_Y\,A^3)=\frac{\Gamma_{\phi\phi\to YY}}{m_\phi}\,\frac{\rho_\phi\,(1+\omega_\phi)}{H}\,A^2\,dA .
 \ee
One should note that $\Gamma^{DM}_{\phi\phi\rightarrow ff} \propto \rho_\phi/m_\phi$, which makes fermion production slower compared to bosonic one during reheating.
However, production of both DMs and radiation are expected to be completed well before the end of reheating. As a result the comoving $(n_Y, \rho_R)$ become constant at reheating end. Therefore, present DM abundance can be safely calculated at the reheating end and is expressed as 
\be \label{DMab}
\Omega_Y\,h^2=\frac{m_Y\,n_Y(A_{re}) A_{re}^3}{\rho_R(A_{re}) A_{re}^4}\,\frac{A_{re}T_{re}}{T_{0}}\,\Omega_R\,h^2\ =0.12,
\ee
$\Omega_R h^2 =4.16\times 10^{-5}$ is the present radiation abundance.
\begin{figure}[t]
	{\includegraphics[width=9cm,height=3.2cm]{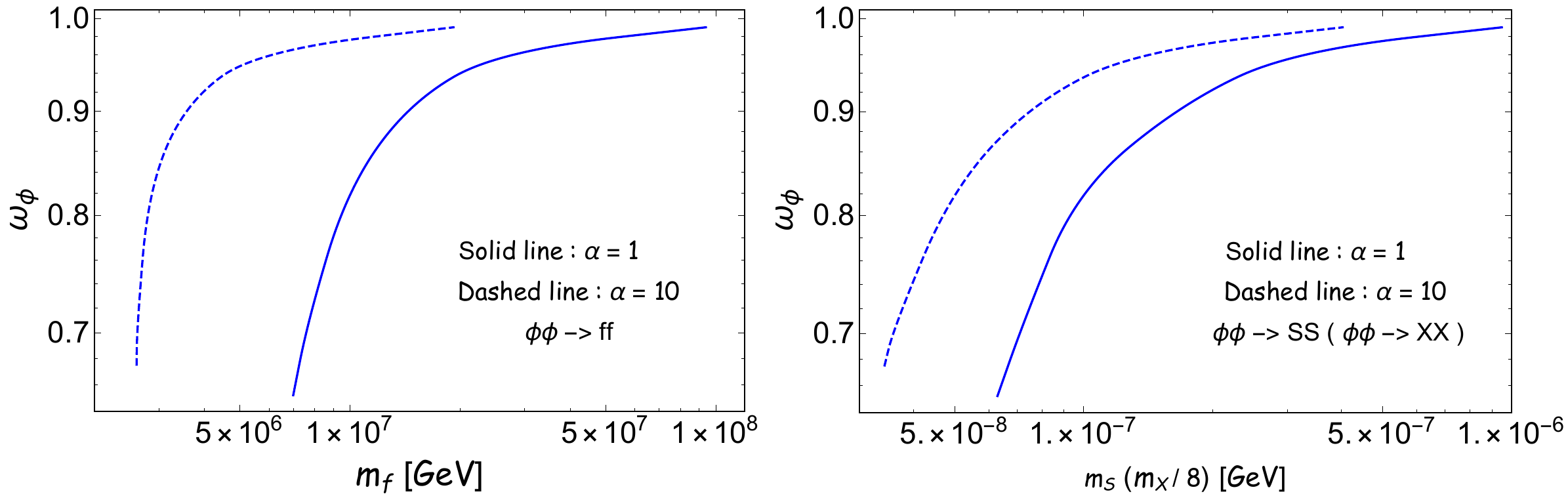}}
	\caption{Variation of $\omega_\phi$ with respect to DM mass.}
	\label{darkmattermass}
\end{figure} 
Upon substitution of the $\Gamma^{DM}_{\phi\phi\rightarrow YY}$ (see, for instance, Eqn.\ref{decay}) into $\ref{DMcomoving}$ and after straightforward integration of Eq.\ref{DMcomoving}, one can find the comoving DM number density, $n_Y^{com} =n_Y A_{re}^3$, at the end of reheating 
for {\it fermion} and scalar/vector DM as,
\begin{eqnarray} \label{darkfermion}
&&n_f^{com}\simeq\frac{3 H_{end}^3}{2048 \pi}\frac{1+\omega_\phi}{1-\omega_\phi}\left(\frac{m_f}{m_\phi^{end}}\right)^2 \left(1-e^{-\frac{3N_{re}}{2}(1-\omega_\phi)}\right),\nno\\
&&n_S^{com} =8 n_X^{com} =\frac{3H_{end}^3\,(1+\omega_\phi)}{512(\pi+3\pi\omega_\phi)} ,
\end{eqnarray}
respectively. Now, using this comoving number densities and the abundance expression (Eq.\ref{DMab}), we can constrain the DM mass. For better visualization, see Fig.\ref{darkmattermass}.\\
\underline{\it Model independent constraints on $m_Y$:}
We have already obtained the model independent constraint on $(H_{end},\omega_{\phi})$ on which  $\Omega_Y\,h^2$ depends explicitly through Eqs.\ref{darkfermion}. Therefore, successful GRE along with the correct DM relic abundance immediately put tight constraints on the allowed mass range for fermionic DM as, $2\times10^5 ~\mbox{GeV} \leq m_f \leq3\times10^8~\mbox{GeV}$, and for scalar/vector DM as $50~\mbox{eV} \leq m_{S}, \gamma m_{X} \leq 1000$ GeV. Origin of higher $m_f$ can be understood from the additional mass suppression $\left(m_Y/m_\phi\right)^2$ in the $\phi\phi\rightarrow ff$ decay width, which suppresses the fermionic DM number density. This requires enhanced value of $m_f$ to satisfy the abundance. \\
\underline{\it Model dependent constraints on $m_Y$:}
Considering $\alpha =(1,10)$ in the Fig.\ref{darkmattermass} we plotted $(m_Y ~\mbox{Vs} ~\omega_{\phi})$ within the allowed range of $\omega_{\phi}$ obtained previously. Important point to realize from the figure that for a specific value of $\omega_{\phi}$ DM mass is unique. The allowed fermionic masses turned out to be within $\{(7\times 10^6,\, 9\times 10^7),\,(3\times 10^6,\, 2\times 10^7)\}$ GeV for $\alpha=(1,\,10)$. For bosonic DM, it is within $\{(60, 1000),(30, 400)\}$ eV for $\alpha=(1,10)$. {\it Therefore, addition to selecting limited class of inflation models successful GRE predicts DM mass $m_Y$ within a very narrow range of values.}\\
\begin{figure}[t]	{\includegraphics[width=8.9cm,height=3.1cm]{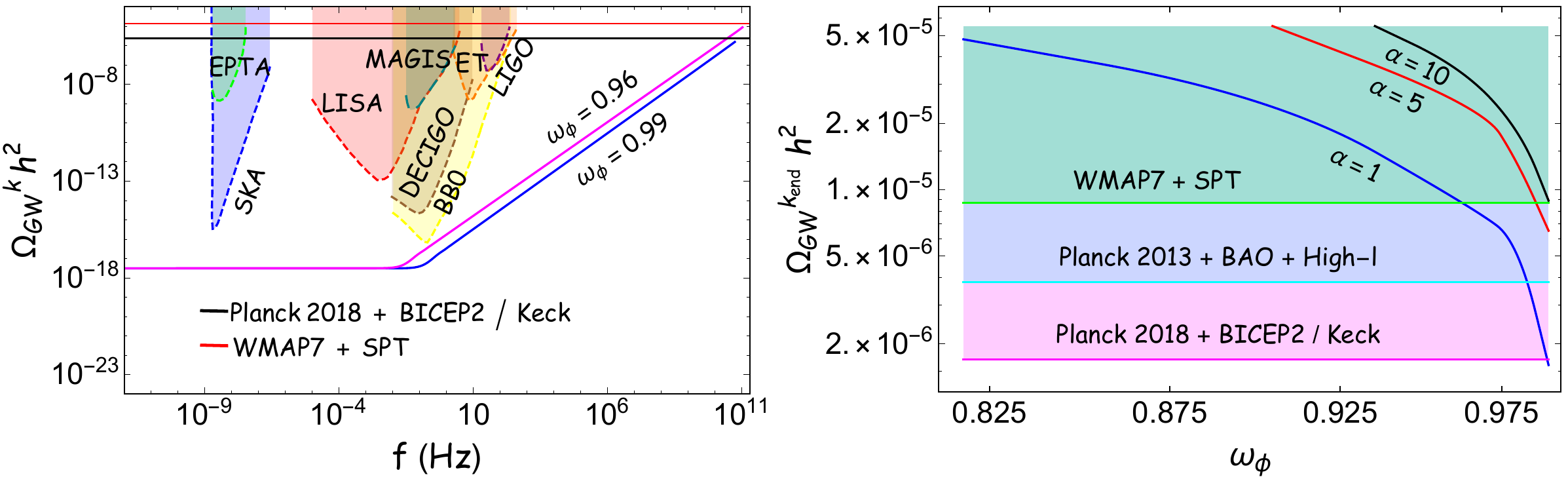}}
	\caption{{\bf Left panel :} Behavior of $\Omega_{GW}^k$ over a wide a range of frequency $f=k/2\pi$ for $\alpha=1$. {\bf Right panel :} $\Omega_{GW}^{k_{end}}$ Vs $\omega_\phi$ for three different values of $\alpha$ and the shaded regions are forbidden from three different BBN bounds.}
	\label{gwbound}
\end{figure}  
{\bf PGWs and constraints:}
PGWs, (see Refs.\cite{Grishchuk:1974ny,Starobinsky:1979ty,Guzzetti:2016mkm,Caprini:2018mtu}) is one of the profound predictions of inflation. 
It plays as a unique probe of the early Universe. Particularly, the evolution of GWs and its amplitude are sensitive to the inflationary energy scale and the post inflationary EoS of the Universe. Extremely weak coupling with matter fields helps PGWs to carry precise information about its origin and subsequent evolution over a large cosmological time scale.
Even though, we have not observed PGWs yet \cite{LIGOScientific:2016jlg,Punturo:2010zz,Crowder:2005nr,Seto:2001qf,LISA:2017pwj,Janssen:2014dka}, simple cosmological upper bound on its strength during BBN will be shown to further tighten the bounds on the parameters discussed above. We focus on the behavior of PGWs spectrum for modes within $ k_{re} < k < k_{end}$ which re-enter the horizon during GRE after inflation. $(k_{re},k_{end})$ re-enter the horizon at the end of inflation and at the end of GRE respectively. Assuming GRE phase is dominated by $\omega_{\phi}$, the PGWs spectrum today is calculated as (see Ref.\cite{Haque:2021dha} for detailed derivation)
\be \label{GW}
\Omega^k_{GW}h^2\simeq \Omega_R h^2\textit{P}_T(k)\frac{4\mu^2}{\pi}\Gamma^2\left(\frac{5+ 3\omega_{\phi}}{2+6\omega_{\phi}} \right)\left(\frac{k}{2\mu k_{re}}\right)^{n_{GW}}
\ee
Where, $\mu=\frac{1}{2}(1+3\omega_\phi)$ and the index of the spectrum, $n_{GW}=-{(2-6\,\omega_\phi)}/{(1+3\omega_\phi)}$. The tensor power spectrum is, $\textit{P}_T(k)=H_{end}^2/12 \pi^2 M_p^2$. To this end we would like to state that for $k < k_{re}$, PGWs spectrum today is $\Omega^k_{GW} (k) h^2  \sim \Omega_Rh^2\,H_{end}^2/12\,\pi^2 M_p^2$, which is scale-invariant for de-Sitter inflation. Eq.\ref{GW} indicates that $\Omega^k_{GW}$ increases with increasing $k$ for $\omega_{\phi}>1/3$ (see Fig.\ref{gwbound}). Effective number of relativistic degrees of freedom during BBN place an upper limit on $\Omega_{GW}^k$ (see Fig.\ref{gwbound}) \cite{Pagano:2015hma}. We will analyze how this upper limit will give even tighter constraints on the parameters.\\
%
\underline{\it Model independent constraints:}
The maximum possible $k= k_{end}$ and the relation $k_{end}/k_{re} =  Exp[N_{re}(1+3\omega_\phi)/2]$ indicate $\Omega_{GW}^{k_{end}} h^2$ being dependent only on $(\omega_{\phi}, H_{end})$, and hence provide further constraints in $(\omega_{\phi}, H_{end})$. Considering constraints on $\Omega_{GW}^{k_{end}} h^2$ within $(1.7\times 10^{-6}, 8.4 \times 10^{-6})$ from different data set (see  Fig.\ref{gwbound}), allowed range of EoS becomes $0.97 < \omega_{\phi} \leq 0.99 $. This is much tighter compared to the constraints from GRE only and increasingly hinting towards the GRE phase being kination domination. This stringent constraint on $\omega_{\phi}$ turned out to be consistent only with the inflationary e-folding number around $N_k \simeq 62$. Constrain on $(H_{re}, T_{re}, m_f)$ remains nearly same as before, but scalar DM mass range further tightens into $(400,1000)$ eV. \\
\underline{\it Model dependent constraints:} First panel of the Fig.\ref{gwbound} suggests, 
if one considers most conservative bound on $\Omega_{GW}^{k_{end}} \,h^2\leq1.7\times 10^{-6}$ obtained from data set Planck-2018 + BICEP2/Keck array \cite{Clarke:2020bil}, $\alpha =1$ with $\omega_{\phi} \simeq 0.99$ appears to be the only allowed model which satisfies all the constraints. However, once relaxing the bound within $(1.7\times 10^{-6} \leq \Omega_{GW}^{k_{end}} \leq 8.4 \times 10^{-6})$ taking into account WMAP7\cite{WMAP:2010qai} and SPT\cite{Keisler:2011aw}, allowed range of $\omega_{\phi}$ get narrowed down within $\{(0.96, 0.99),(0.986,0.990)\}$ for $\alpha=(1,5)$ accordingly. Whereas, any $\alpha > 10$ are completely excluded.
Within the allowed value of $\alpha=(1,10))$, maximum allowed range of scalar spectral index $n_s$ becomes $(0.9671, 0.9683)$, reheating temperature $T_{re}$ becomes ($2\times10^5, 2\times10^6$) GeV, fermionic DM mass becomes $(10^7, 10^8)$ GeV, and scalar/vector DM mass becomes ($400,1000$) eV. However, all these ranges actually shrink towards their lower value as one goes from $\alpha = 1 \to 10$.

 \begin{figure}[t]
{\includegraphics[width=8.5cm,height=2.9cm]{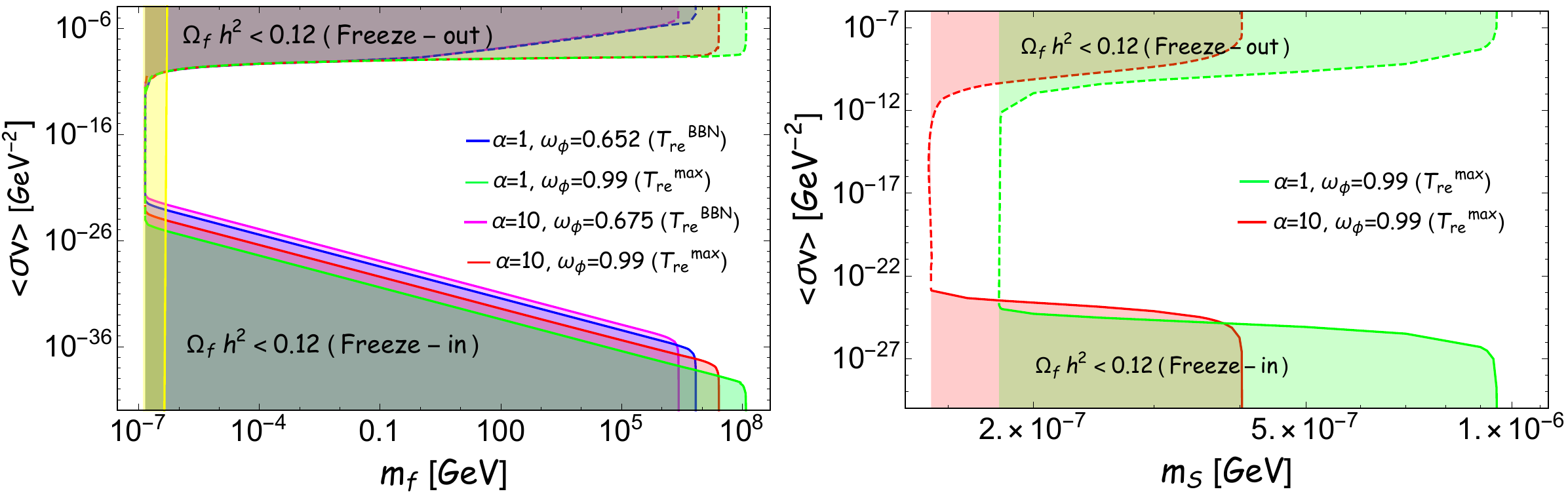}}

	\caption{ Plot of $\langle \sigma v \rangle$ Vs $m_Y$ of fermionic (left panel) and scalar (right panel) DM.  
	 The Yellow band corresponds to minimum DM mass bound taken from \cite{Alvey:2020xsk}.}
	\label{darkmattermasssigmav}
\end{figure} 
{\bf Conclusions:} GRE appeared to be a minimal production mechanism scenario through which our present state of the Universe can be obtained after inflation. 
Being DM mass is the only free parameter in the DM sector, successful GRE puts stringent constraints on possible DM mass and may pave the way toward constructing DM models. The scenario further restricts inflationary model parameters once we project our result in the $n_s-r$ plane with the latest available combined data from BICEP/Keck and Planck. Considering available bounds on PGWs spectrum and DM abundance, GRE selects those inflation models with a unique value of $N_k\sim 62$ and inflaton EoS above $>0.97$ during reheating. Consequently reheating temperature must be $T_{re} < 10^8$ GeV, fermionic DM mass should lie within $2\times10^5 ~\mbox{GeV} < m_f <3\times10^8~\mbox{GeV}$, and scalar/vector DM mass within $400~\mbox{eV} < m_{S}, \gamma m_{X} <1000$ eV.
The results just mentioned above are obtained without specifying any model except the generic de Sitter type inflation. However, if we consider a specific model such as $\alpha$-attractor, more narrower bounds are obtained due to its small prediction of $r$. The upper limit on $\alpha$ strictly bounded by $\alpha\leq15$ from the combined data of BICEP/Keck and Planck with GRE as a reheating background.

 To this end, let us point out that if we take into account the modified decay widths properly accounting for the oscillating inflaton zero-mode \cite{Clery:2021bwz}, all our predictions remain quantitatively the same except the fermionic DM mass range shifted towards the lower value by one order.\\  
So far, all GRE predictions seem to be independent of any new physics in the radiation sector. We have also shown where our GRE scenario lies if we consider different non-gravitational couplings between inflaton and radiation sector(see Fig.\ref{bound}). Moreover, if DM sector couples directly with the radiation bath with thermally averaged cross-section times velocity $\langle \sigma v\rangle$, then the DM masses obtained previously transformed into maximum one $m_Y^{max}$ in $(\langle \sigma v\rangle, m_Y)$ space   \cite{Haque:2021mab} (see Fig.\ref{darkmattermasssigmav}). Upon decreasing DM mass, to our surprise, the existence of nearly model-independent minimum DM mass $m^{min}_Y$ is observed where freeze-in and freeze-out mechanisms meet together. Such observation was also never reported before in the literature. This phenomenon is expected as decreasing $m_Y$ requires increasing $\langle \sigma v\rangle$ during freeze-in, and at its threshold value $m^{min}_Y$ the DM thermalizes with radiation bath where freeze-out begins. The value of $m^{min}_{Y}$ turned out as $\sim 150$ eV for fermion DM irrespective model parameters. However, for fermionic DM, the most compact DM-dominated object called dwarf spheroidal galaxies are known to provide the lowest bound (Tremaine-Gunn (TG) bound) on its mass $m_f\geq 590$ eV at $68\%$ CL 
	\cite{Alvey:2020xsk} shown in yellow shaded region.
	Finally, we want to point again that $m_Y^{max}$ is set to be the maximum possible DM mass for both freeze-in ($\langle \sigma v\rangle\to 0$) and freeze-out ($\langle \sigma v\rangle \to\infty$) scenarios if one satisfies the present DM abundance. Therefore, if DM with $m_Y  > m_Y^{max}$ is detected, it will rule out the possibility of purely gravitational reheating after inflation.\\  
{\bf Acknowledgments:}
M.R.H wish to acknowledge support from the Science and Engineering 
Research Board~(SERB), Government of India~(GoI), for the SERB National Post-Doctoral fellowship, File Number: PDF/2022/002988. D.M wish to acknowledge support from the Science and Engineering 
Research Board~(SERB), Department of Science and Technology~(DST), GoI,
through the Core Research Grant CRG/2020/003664. We like to thank the HEP and Gravity groups at IIT Guwahati
for useful discussions.

\end{document}